\documentclass{IEEEtran4PSCC}

\usepackage{graphicx}

\usepackage[cmex10]{amsmath}
\makeatletter
\let\old@ps@headings\ps@headings
\let\old@ps@IEEEtitlepagestyle\ps@IEEEtitlepagestyle
\def\psccfooter#1{%
    \def\ps@headings{%
        \old@ps@headings%
        \def\@oddfoot{\strut\hfill#1\hfill\strut}%
        \def\@evenfoot{\strut\hfill#1\hfill\strut}%
    }%
    \def\ps@IEEEtitlepagestyle{%
        \old@ps@IEEEtitlepagestyle%
        \def\@oddfoot{\strut\hfill#1\hfill\strut}%
        \def\@evenfoot{\strut\hfill#1\hfill\strut}%
    }%
    \ps@headings%
}
\makeatother

\psccfooter{%
        \parbox{\textwidth}{\hrulefill \\ \small{24th Power Systems Computation Conference} \hfill \begin{minipage}{0.2\textwidth}\centering \vspace*{4pt} \includegraphics[scale=0.06]{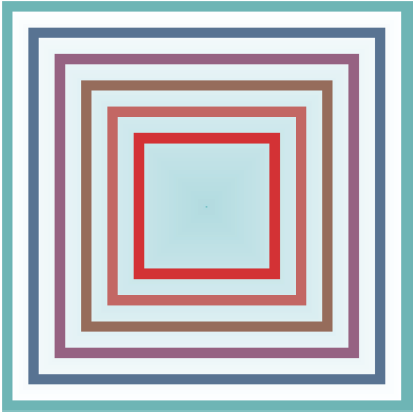}\\\small{PSCC 2026} \end{minipage} \hfill \small{Limassol, Cyprus --- June 8-12, 2026}}%
}
\usepackage[utf8]{inputenc}       
\usepackage{inconsolata}          

\usepackage{amsmath}              
\usepackage{amsthm}               
\usepackage{amsfonts}             
\usepackage{mathtools}            
\usepackage{breqn}                

\usepackage{algorithm}            
\usepackage{algpseudocode}

\usepackage{tikz}                 
\usetikzlibrary{shapes,arrows}    
\usepackage{circuitikz}           
\usepackage{subcaption}

\usepackage{multirow}             
\usepackage{booktabs}             
\usepackage{array}
\usepackage{makecell}
\usepackage{tabularx}

\usepackage{gensymb}              
\usepackage{lipsum}               
\usepackage{matlab-prettifier}    
\usepackage{xurl}                  
\usepackage[hidelinks]{hyperref}

\usepackage[hidelinks]{hyperref}  
\usepackage{cleveref}             
\usepackage{cite}                 
\usepackage{balance}

\newtheorem{problem}{Problem}

\begin{document}
%

\title{Offset-free Data-Driven Predictive Control for Grid- Connected Power Converters in Weak Grid Faults}


\author{
\IEEEauthorblockN{Ivo Kraayeveld\\Thomas de Jong\\Mircea Lazar}
\IEEEauthorblockA{Department of Electrical Engineering, Control Systems\\Eindhoven University of Technology\\
Eindhoven, The Netherlands\\
\{i.t.kraayeveld;t.o.d.jong;m.lazar\}@\{student.tue;tue;tue\}.nl}
}

\maketitle

\begin{abstract}
    Grid-connected power converters encounter significant stability challenges during weak grid faults, when conventional PI-based controllers exhibit an oscillatory response and poor fault-ride-through performance. This paper addresses this problem by replacing the conventional outer PI controllers that regulate DC-link and PCC voltages with an offset-free data-driven predictive controller. The developed algorithm leverages either pre-fault or fault-time data to construct input-output predictors, yielding offset-free control without the need for physics-based modelling. Simulation results show that pre-fault offset-free DPC doubles the critical equivalent grid impedance that can be handled and reduces the root mean squared error during faults by a factor of 40, while maintaining computation times comparable to conventional PI control. These findings demonstrate that the developed offset-free data predictive controller offers a simple, robust, and computationally efficient alternative to conventional control, significantly enhancing fault-ride-through capabilities of converters in weak grids.

\end{abstract}

\begin{IEEEkeywords}
    Data-driven predictive control, grid-connected power converters, offset-free tracking, grid fault-ride-through
\end{IEEEkeywords}

\thanksto{\noindent Submitted to the 24th Power Systems Computation Conference (PSCC 2026).}

\section{Introduction}
\begin{figure}
        \centering
        \includegraphics[width=\columnwidth]{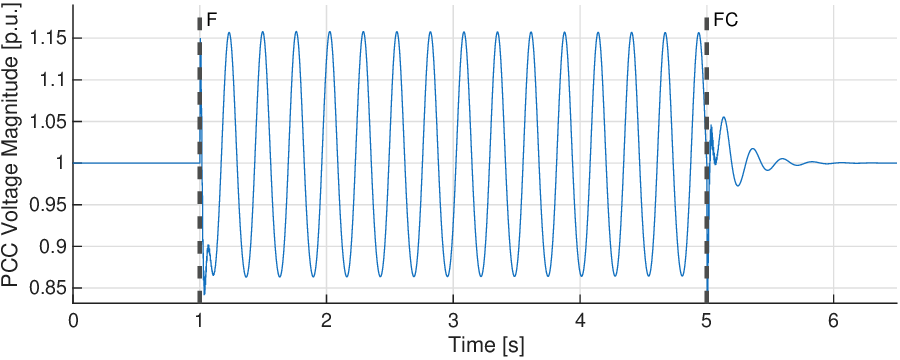}
        \caption{Simulated PCC voltage of a GCPC with a critically weak grid fault. The lines labelled F and FC indicate the start of the fault and the moment it is cleared, respectively.}
        \label{fig:GCPC Conventional Control}
    \end{figure}

    Over the last decades, the penetration of renewables in global energy generation has grown rapidly. The trend of replacing synchronous resources with converter-based resources (CBRs) brings forth challenges like sub-synchronous oscillations (SSOs) in grid-connected power converters (GCPCs). Practical examples of such problematic behaviour included the faults in the UK~\cite{Ofgem20209thReport} and Spain~\cite{Lombardi2025MiscalculationFinds}. In both cases, GCPC controllers interacted with a weak grid during a voltage fluctuation, with oscillations as a result. The decreasing number of synchronous machines leads to a shortage of reactive power during faults, slowing down current reference tracking and, therefore, voltage control. This effect, similar to an increased equivalent grid impedance, leads to over-aggressive control action from the conventional, PI-based controllers in grid-following (GFL) schemes. The result of this can be seen in Fig.~\ref{fig:GCPC Conventional Control}, where the point of common coupling (PCC) voltage shows sustained oscillations due to a fault, meaning poor fault-ride-through performance.

    Although grid-forming (GFM) control, such as virtual synchronous generators or virtual oscillator control~\cite{Chen2018ModellingSystem, Lu2022BenchmarkingControl}, could mitigate this, it would require completely replacing existing GFL control implementations. Instead, existing GFL GCPC control could be improved to increase robustness through more advanced control principles. An overview of different methods can be found in~\cite{Babayomi2025AdvancedTrends}. Model predictive control (MPC) has recently emerged as an attractive alternative to classical control for inverters and modular multi-level converters, see, e.g.,~\cite{Geyer2016ModelDrives,Shetgaonkar2023ModelSystems,ReyesDreke2022Long-HorizonConverters}. Moreover,~\cite{Merabet2018RobustCapability,Mahr2021AdvancedController} demonstrated that MPC enables CBRs to provide grid-fault ride-through capability, i.e., maintaining grid connection during faults while actively contributing to grid stability. One of the main issues of MPC, however, is its inherent dependence on an accurate prediction model. Obtaining an accurate model is particularly difficult in modern power systems, where dynamics are time-varying and different components are used.
    
    To address this, data-driven predictive control (DPC) methods emerged as a popular alternative to MPC. Unlike traditional MPC, DPC builds input-output predictors directly from data, thereby eliminating the need for a physics-based model and state observer~\cite{Verheijen2023HandbookDesign}. Recently, several works have demonstrated the potential of DPC in addressing key challenges of modern power systems. For instance,~\cite{Bilgic2022TowardSystems} applies DPC to the operation of distributed multi-energy systems, while~\cite{Zhao2024Data-drivenInertia} introduces a data-driven adaptive predictive frequency control strategy for multi-area power systems with unknown and time-varying inertia. Similarly,~\cite{Zhao2024DirectOptimization} proposes an adaptive data-driven controller for stabilising power converters. The applicability of DPC in real-world power systems has also been demonstrated in~\cite{graf2025gridconnecteddatadriveninvertercontrol}, where the conventional outer control layer was replaced by a DPC scheme, transient predictive control, and experimentally validated on an inverter connected to the Munich grid. Very recently,~\cite{Leng2025DeePConverter:Converters} has designed a data-enabled predictive controller with integral action (iDeePC) for GCPCs and has shown that iDeePC outperforms conventional control (CC) methods. Compared to other DPC algorithms~\cite{Verheijen2023HandbookDesign}, however, iDeePC complexity increases with the dataset size, which limits the usable data amount. 
    
    These examples show the growing maturity of DPC methods and their potential to improve GCPC control. Therefore, this paper presents an implementation of integral subspace predictive control (iSPC)~\cite{Lazar2022Offset-freeControl}, an alternative offset-free DPC algorithm that converges to the same predictor as iDeePC, for GCPCs in weak grid faults. In the developed architecture, iSPC replaces the conventional outer layer controlling the DC-link and PCC voltage magnitudes. To validate the improvement brought by iSPC, the critical grid impedance and voltage drop that can be handled by CC are first identified. Then it is shown that iSPC can handle double the impedance for a similar grid voltage drop. Compared to iDeePC, the developed iSPC solution can handle large datasets, e.g. 10,000 DC-link and PCC voltage magnitude measurements. Regarding real-time implementation, an analytical iSPC solution can be computed with similar computational complexity as conventional PI control.

    The main contributions are: (i) reproduction of oscillations that occur during weak grid faults for GCPC with conventional PI-controllers using a realistic grid model; (ii) implementation of iSPC for GCPC control; and (iii) validation of iSPC in critical faults and comparison with PI-controllers.

    \subsubsection*{Notation and basic definitions}
        In this paper, it is assumed that the system is balanced, meaning signals are given in their single-phase representation using the Park-Clarke transform, e.g. for the grid current, given a rotating reference frame $f$:
        \begin{equation}
            \vec{I}_g(t) = \vec{I}_{g}^{dq|f}(t)e^{j\theta_f(t)}=\left(i_g^{d|f}(t)+j\,i_g^{q|f}(t)\right)e^{j\theta_f(t)},
            \label{eq:dq-representation}
        \end{equation}
        where $\vec{I}_{g}^{dq|f}(t)$ is the complex value produced by the projection, which can also be represented in vector notation as
        \begin{equation}
            I_g^{dq|f}(t) = \begin{bmatrix}
                i_g^{d|f}(t)\\
                i_g^{q|f}(t)
            \end{bmatrix}.
        \end{equation}
        Time dependence is omitted in what follows, unless required in the discrete-time setting. Furthermore, $\text{col}(\xi_1, \dots, \xi_q) := \begin{bmatrix} \xi_1^\top & \dots & \xi_q^\top \end{bmatrix}^\top$ is defined. Given a matrix $A\in \mathbb{R}^{m\times n}$, its generalised pseudo-inverse is denoted as $A^\dagger$. Given a signal $v:\mathbb{N}\rightarrow \mathbb{R}^{n_v}$, $k\geq 0$ and $j\geq k+1$, $\bar{\mathbf{v}}(k,j):=\text{col}(v(k),\dots,v(k+j-1))$. The identity matrix of dimension $n$ is denoted as $\textbf{I}_n$, and $\textbf{O}_{n\times m}$ denotes a matrix with $n$ rows and $m$ columns of zeros. Dimension subscripts are omitted when clear from context.

\section{Preliminaries}
\begin{figure}
        \centering
        \includegraphics[width=\columnwidth]{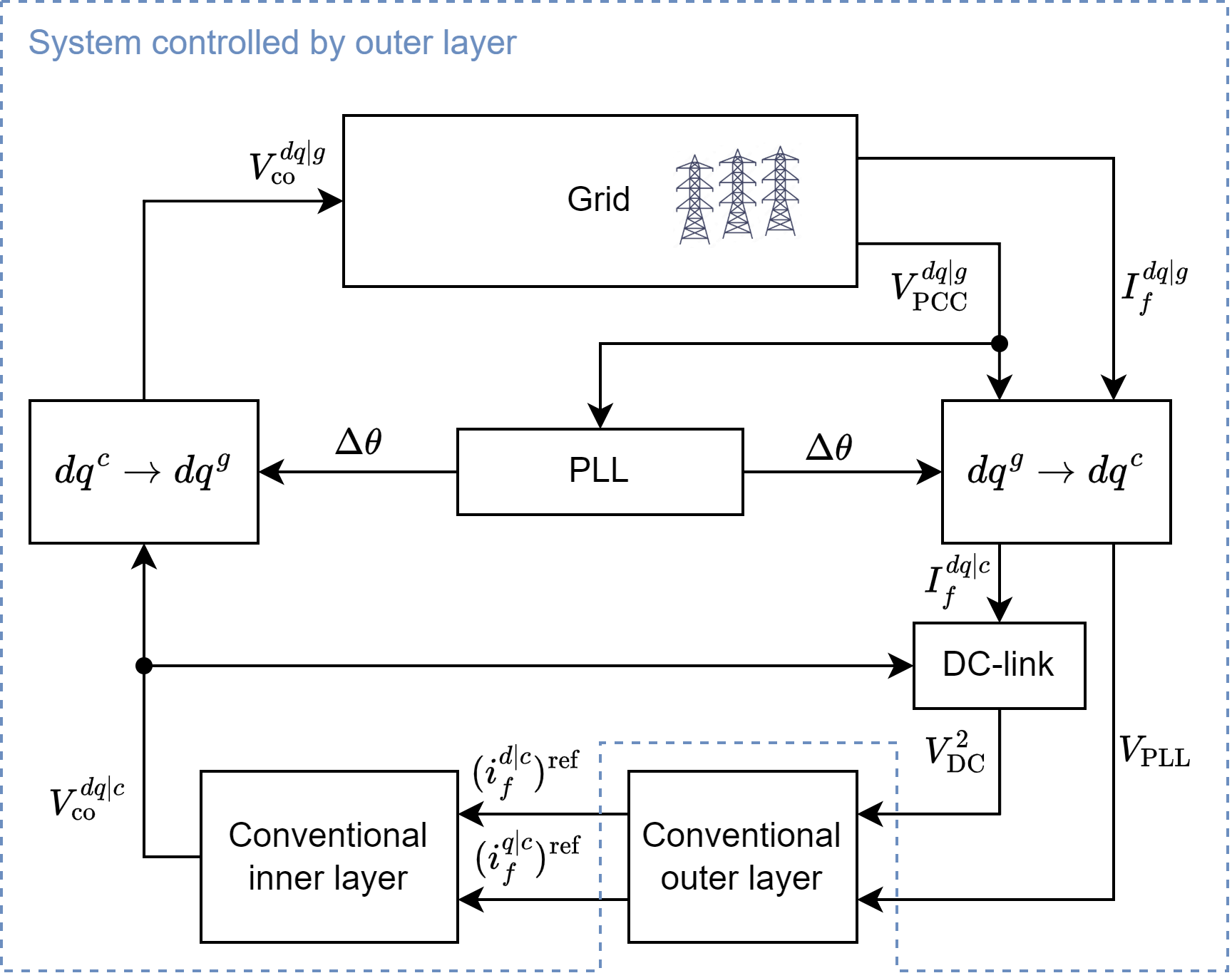}
        \caption{Overview of different system components for a GCPC using CC.}
        \label{fig:conv-control-overview}
    \end{figure}

    The continuous-time grid model used in this paper is similar to the one used in earlier research on SSOs due to controller-weak grid interactions~\cite{Li2019StabilityGrids}. The overview of the system with CC can be found in Fig.~\ref{fig:conv-control-overview}.
    
\begin{figure}
        \centering
        \vspace{-1.5em}
        \begin{subfigure}{\columnwidth}
            \centering
            \begin{circuitikz} [scale=.60/.8, transform shape]
                \draw (0,0) node[ground] {};
                \draw (0,0) to[V, l_=$\vec{V}_{\text{co}}$] ++(0,1)
                to[short] ++(0,0.37)
                to[short] ++(0.4,0)
                to[R, l=$R_{f}$] ++(1.18,0)
                to[short] ++(0.41,0)
                to[L, l=$L_{f}$] ++(0.9,0)
                to[short, i=$\vec{I}_f$] ++(1.4,0)
                ++(0,1) node{$\vec{V}_{\text{PCC}}$}
                ++(0,-1)
                to[short] ++(0,-0.54)
                to[C, l=$C_f$] ++(0,-0.54)
                to[short] ++(0,-0.27)
                node[ground] {}
                ++(0,1.35)
                to[short] ++(0.3,0)
                to[short, i=$\vec{I}_g$] ++(.64,0)
                to[short] ++(0.3,0)
                to[R, l=$R_{\text{ca}}$] ++(1.18,0)
                to[short] ++(0.41,0)
                to[L, l=$L_{\text{ca}}$] ++(0.9,0)
                to[short] ++(0.3,0)
                coordinate (PCC)
                to[short] ++(0.3,0)
                to[L, l=$L_{g}$] ++(0.9,0)
                to[short] ++(0.58,0)
                to[short] ++(0,-0.27)
                to[V, l_=$\vec{V}_{g}$] ++(0,-1.08)
                node[ground] {};
            \end{circuitikz}
            \subcaption{Grid model}
            \label{fig:SinglePhaseEqGrid}
            \vspace{1em}
        \end{subfigure}
        \begin{subfigure}{\columnwidth}
            \centering
            \includegraphics[width=\columnwidth]{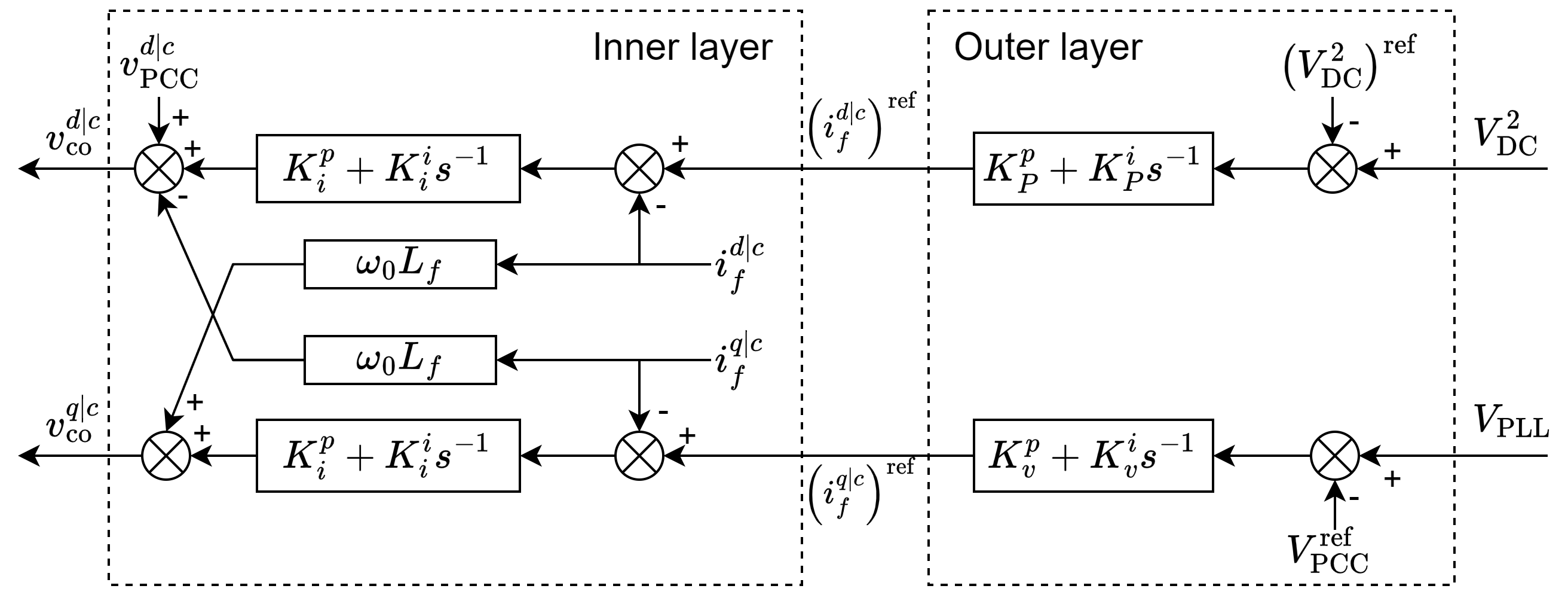}
            \subcaption{Conventional control}
            \label{fig:PIControlBlock}
        \end{subfigure}
        \caption{Converter modelling and control components: (a) a converter connected through a filter and transmission cable to the grid, (b) a block scheme of the conventional PI-control.}
        \label{fig:ConverterSystem}
    \end{figure}
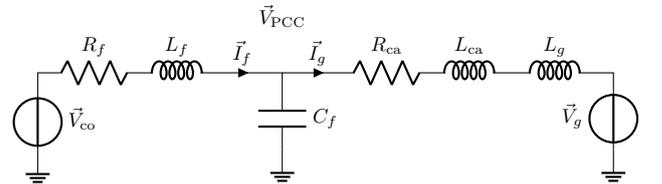
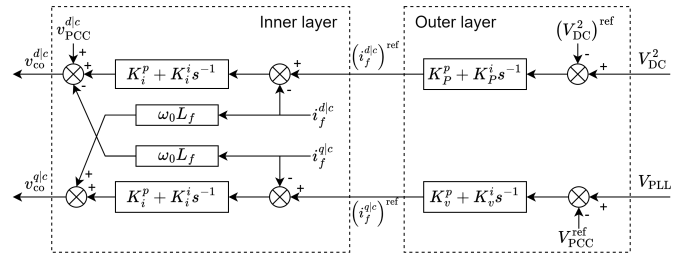

    The single-phase equivalent circuit can be found in Fig.~\ref{fig:SinglePhaseEqGrid}. It shows the converter as an ideal voltage source, followed by an LC filter and a transmission cable. The grid is modelled with a Thevenin equivalent impedance and voltage source, as commonly used in literature studying grid behaviour~\cite{Hu2021PowerControl}. The equivalent impedance to the stiff grid source is a representation based on the short-circuit ratio, a measure of the grid strength during a fault.
    
    The dq-representation reference frames are either the grid frame $g$, which has the d-axis aligned to the grid voltage vector, or the converter frame $c$, which tracks the PCC voltage vector with its d-axis. To track this vector, a phase-locked loop (PLL) is used as in~\cite{Li2019StabilityGrids}. The PLL is nonlinear due to the trigonometric functions in the frame conversion.
    
    Finally, the dynamics of the DC-link capacitor are given by
    \begin{equation}
        \frac{C_{\text{DC}}(V_{\text{DC}}^{2})^{\text{base}}}{2P^{\text{base}}}\frac{dV_{\text{DC}}^2}{dt} = P_{\text{wind}}-P,
        \label{eq:dc-link-dynamics}
    \end{equation}
    where $V_{\text{DC}}^2$, $P_{\text{wind}}$ and $P$ are the DC-link voltage squared, the power flowing in from the generation side of the power converter and the power flowing out of the converter, respectively. All three are expressed in p.u., $P_{\text{wind}}$ is considered constant and $P=v_{\text{co}}^{d|c}i_{f}^{d|c}+v_{\text{co}}^{q|c}i_{f}^{q|c}$. $P^{\text{base}}$ and $(V_{\text{DC}}^2)^{\text{base}}$ are base quantities, as used in the per unit system. 
        
    \subsection{Conventional control scheme}
        CC aims to control the DC-link voltage and PCC voltage magnitude, with the latter approximated with the d-component of the PCC voltage in the converter frame, provided by the PLL, i.e. $V_{\text{PLL}}$~\cite{Examples2020PES-SSOExamples}. The controller structure is shown in Fig~\ref{fig:PIControlBlock}. The figure shows an outer layer with SISO-based PI blocks, which regulate the DC-link voltage and the PCC voltage magnitude, at a slower time scale. The equations for the outer control layer are
        \begin{equation}
            \begin{split}
                (i_f^{d|c})^{\text{ref}} &= \left(V_{\text{DC}}^2-(V_{\text{DC}}^2)^{\text{ref}}\right)(K_P^p+K_P^is^{-1}),\\
                (i_f^{q|c})^{\text{ref}} &= \left(V_{\text{PLL}}-V_{\text{PCC}}^{\text{ref}}\right)(K_v^p+K_v^is^{-1}).
            \end{split}
        \end{equation}
        The inner layer for reference current tracking is defined as
        \begin{equation}
            \scalebox{0.95}{$
            \begin{aligned}
                v_{\text{co}}^{d|c} &= \left(i_f^{d|c}-(i_f^{d|c})^{\text{ref}}\right)(K_i^p+K_i^is^{-1})-\omega_0\tilde{L}\,i_f^{q|c}+V_{\text{PLL}},\\
                v_{\text{co}}^{q|c} &= \left(i_f^{q|c}-(i_f^{q|c})^{\text{ref}}\right)(K_i^p+K_i^is^{-1})+\omega_0\tilde{L}\,i_f^{d|c}.
            \end{aligned}$}
        \end{equation}
        CC gains are tuned to nominal conditions, i.e. negligible grid impedance. During weak-grid faults, this no longer holds, slowing the response. Lacking derivative terms, the controller acts more aggressively to prolonged error, reducing phase margin. This shifts the complex poles further to the right-hand plane~\cite{Fan2018AnInterconnection}, reducing stability as shown in Fig.~\ref{fig:GCPC Conventional Control}. To mitigate this issue, the outer PI controllers, in this paper, are replaced with a single iSPC controller, which is described next.

\section{Integral subspace predictive control}
    To implement iSPC with minimal modifications of the overall CC scheme, the same inputs and outputs as the conventional outer layer will be used. These are defined as
    \begin{equation}
        \begin{matrix}
            u(k)=\begin{bmatrix}
                (i_f^{d|c})^{\text{ref}}(k)\\(i_f^{q|c})^{\text{ref}}(k)
            \end{bmatrix}, &y(k)=\begin{bmatrix}V_{\text{DC}}^{2}(k)\\V_{\text{PLL}}(k)\end{bmatrix},
        \end{matrix}
    \end{equation}
    giving dimensions $n_u=n_y=2$. The iSPC receives sampled output measurements, and the control actions are implemented using a zero-order-hold (ZOH) mechanism, as typically done in digital control. The system is assumed linear around the operating point, i.e.,
    \begin{equation}
        \begin{split}
            x(k+1) &= A\,x(k) + B\,u(k), \,\,\,\,\,\,\, k\in \mathbb{N},\\
            y(k) &= Cx(k),
        \end{split}
    \end{equation}
    where $x(k)\in\mathbb{R}^{n_x}$ is the state and $n_x\in\mathbb{N}$ is the state dimension. This model can be lifted to an integral form as follows
    \begin{equation}
        \begin{split}
            \begin{bmatrix}
                \Delta x(k+1)\\
                y(k+1)
            \end{bmatrix}&=\begin{bmatrix}
                A&\mathbf{O}\\
                CA&\mathbf{I}
            \end{bmatrix}\begin{bmatrix}
                \Delta x(k)\\
                y(k)
            \end{bmatrix}+\begin{bmatrix}
                B\\CB
            \end{bmatrix}\Delta u(k),\\
            y(k)&=\begin{bmatrix}
                \mathbf{O} & \mathbf{I}
            \end{bmatrix}\begin{bmatrix}
                \Delta x(k)\\
                y(k)
            \end{bmatrix},
        \end{split}
    \end{equation}
    where $\Delta x(k):=x(k)-x(k-1)$ and $\Delta u(k) := u(k)-u(k-1)$. The predicted future sequences of inputs and outputs for prediction horizon $N\in\mathbb{N}$ are defined as
    \begin{equation}
        \begin{split}
            \Delta \mathbf{u}(k) &:= \text{col}\left(\Delta u_{0|k},\dots,\Delta u_{N-1|k}\right),\\
            \mathbf{y}(k) &:= \text{col}\left(y_{1|k},\dots,y_{N|k}\right),
        \end{split}
        \label{eq:io-trajectories}
    \end{equation}
    where $i|k$ denotes the prediction for time $k+i$ made at $k$. As mentioned in~\cite{Lazar2022Offset-freeControl}, the future sequence for integral MPC (iMPC) can be expressed through a linear combination of initial conditions and the input sequence as
    \begin{equation}
        \mathbf{y}^{\text{iMPC}}(k) = \Phi_I \begin{bmatrix}
                \Delta x(k)\\
                y(k)
            \end{bmatrix} + \Gamma_{I}\,\Delta\mathbf{u}(k).
    \end{equation}
    The problem with this approach is that it requires state information via $\Delta x(k)$, which the controller cannot access. 
    
    In iSPC\cite{Lazar2022Offset-freeControl}, the multi-step predictor is defined as
    \begin{equation}
        \mathbf{y}^{\text{iSPC}}(k) = \begin{bmatrix} 
            P_{1,I} & P_{2,I} 
        \end{bmatrix}\begin{bmatrix}
            \Delta\mathbf{u}_{\text{ini}}(k)\\
            \mathbf{y}_{\text{ini}}(k)
        \end{bmatrix} + \Gamma^I \Delta\mathbf{u}(k),
    \end{equation}
    which only requires input-output data. Given the initial horizon length $T_{\text{ini}}\in \mathbb{N}$, the input and output vectors are defined as
    \begin{equation}
        \begin{split}
            \Delta \mathbf{u}_{\text{ini}}(k) &:= \text{col}\left(\Delta u(k-T_{\text{ini}}),\dots,\Delta u(k-1)\right),\\
            \mathbf{y}_{\text{ini}}(k) &:= \text{col}\left(y(k-T_{\text{ini}}+1),\dots,y(k)\right).
        \end{split}
        \label{eq:initial-io-trajectories}
    \end{equation}
    For the remainder of the paper, the superscripts for $\mathbf{y}^{\text{iSPC}}$ and $\Gamma^{\text{iSPC}}$ are omitted. To build the iSPC predictor, given the total number of recorded datapoints $T\in \mathbb{N}$, Hankel matrices are defined as
    \begin{equation}
        \begin{split}
            \Delta\mathbf{U}_p &:= \begin{bmatrix} \Delta\mathbf{\bar{u}}(0,T_{\text{ini}}) & \cdots & \Delta\mathbf{\bar{u}}(T-1,T_{\text{ini}}) \end{bmatrix},\\
            \mathbf{Y}_p &:= \begin{bmatrix} \mathbf{\bar{y}}(0,T_{\text{ini}}) & \cdots & \mathbf{\bar{y}}(T,T_{\text{ini}}) \end{bmatrix},\\
            \Delta\mathbf{U}_f &:= \begin{bmatrix} \Delta\mathbf{\bar{u}}(T_{\text{ini}},N) & \cdots & \Delta\mathbf{\bar{u}}(T_{\text{ini}}+T-1,N) \end{bmatrix},\\
            \mathbf{Y}_f &:= \begin{bmatrix} \mathbf{\bar{y}}(T_{\text{ini}}+1,N) & \cdots & \mathbf{\bar{y}}(T_{\text{ini}}+T,N) \end{bmatrix}.
        \end{split}
        \label{eq:hankel-matrices}
    \end{equation}
    The iSPC prediction matrices are then obtained as a solution to the least-squares problem as follows
    \begin{equation}
        \Theta^* = \begin{bmatrix}
                P_{1,I}&P_{2,I}&\Gamma
        \end{bmatrix}=\mathbf{Y}_f\begin{bmatrix}
            \Delta\mathbf{U}_p\\ \mathbf{Y}_p\\ \Delta\mathbf{U}_f
        \end{bmatrix}^{\dagger}.
        \label{eq:theta-estimation}
    \end{equation}
    Computing the pseudo-inverse in~\eqref{eq:theta-estimation} requires that the Hankel data matrix is full-rank, which, in turn, requires a sufficiently persistently exciting input signal. For details, see~\cite{Verheijen2023HandbookDesign}. To achieve this, additive white noise on top of the reference currents was used. 
    
\begin{figure}
        \centering
        \includegraphics[width=.85\columnwidth]{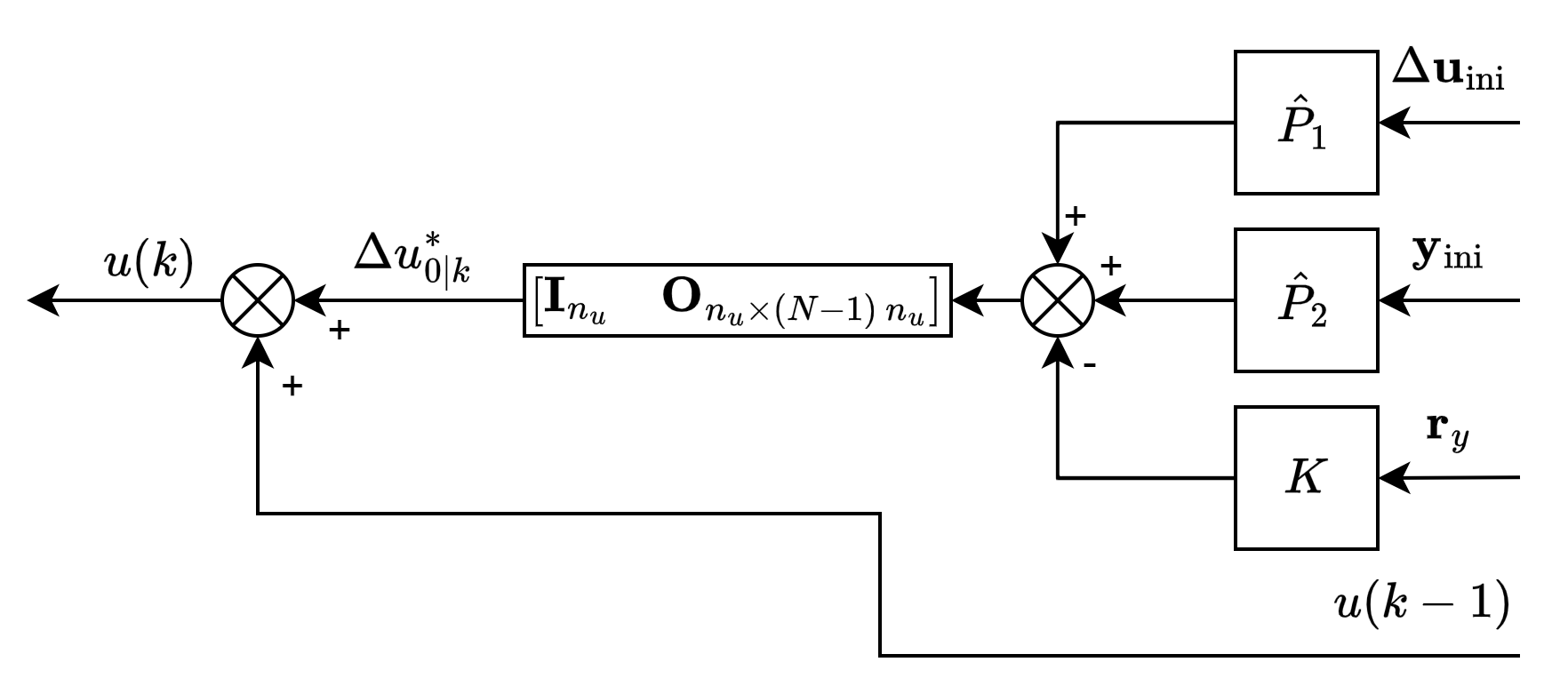}
        \caption{Block scheme of an iSPC controller}
        \label{fig:iDPC-Control-phase-controller}
    \end{figure}

    The iSPC optimisation problem is stated next.
    \begin{problem}[iSPC Problem at time $t=kT_s$]\label{prob:iSPC }
        \begin{subequations}
            \begin{align}
                \min_{\Delta\mathbf{u}(k), \mathbf{y}(k)} \quad 
                & \begin{aligned}[t]
                    & \sum_{i=0}^{N-1} l\big(\Delta u_{i|k}, y_{i|k}, r_y\big)+ l_N\big(y_{N|k}, r_y\big)
                  \end{aligned} \label{prob:cost} \\
                \textnormal{s.t.} \quad 
                & \resizebox{.65\columnwidth}{!}{$
                \mathbf{y}(k) =
                    \begin{bmatrix} P_{1,I} & P_{2,I} \end{bmatrix}
                    \begin{bmatrix}
                        \Delta\mathbf{u}_{\text{ini}}(k)\\
                        \mathbf{y}_{\text{ini}}(k)
                    \end{bmatrix}
                    + \Gamma \Delta\mathbf{u}(k),
                $} \label{prob:dynamics}
            \end{align}
        \end{subequations}
    \end{problem}
    \noindent where $r_y$ is the constant output reference, and the cost functions are defined as
    \begin{equation}
        \begin{split}
            l\big(\Delta u_{i|k}, y_{i|k}, r_y\big)= &\left(y_{i|k}-r_y\right)^\top Q\left(y_{i|k}-r_y\right)\\
            &+\Delta u_{i|k}^\top R\Delta u_{i|k},\\
            l_N\big(\Delta u_{i|k}, y_{i|k}, r_y\big)= &\left(y_{i|k}-r_y\right)^\top P\left(y_{i|k}-r_y\right).
        \end{split}
    \end{equation}
    As shown next, an analytic solution to Problem~\ref{prob:iSPC } can be found (alternatively, input and output constraints could be added, and a numerical solver can solve the problem online). Given the two matrices $\Omega:=\text{diag}(Q,\dots,Q,P)$ and $\Psi:=\text{diag}(R,\dots,R)$, and the vector $\mathbf{r}_y(k)=\text{col}(r_y(k), \dots, r_y(k))$ as the stacked future references, the optimal iSPC control law is
    \begin{equation}
        \begin{split}
            \Delta\mathbf{u}^*(k)=\underbrace{-(\Psi+\Gamma^\top \Omega\Gamma)^{-1}\Gamma^\top \Omega}_{K}(P_{1,I} \Delta \mathbf{u}_{\text{ini}}(k)\\
            +P_{2,I} \mathbf{y}_{\text{ini}}(k)-\mathbf{r}_y(k)).
        \end{split}
        \label{eq:optimal-input-sequence}
    \end{equation}
    Here, $\Delta \mathbf{u}^*(k)$ is the predicted optimal input sequence based on the data. The first input increment $\Delta u^*_{0|k}$ is then used to calculate the input applied to the system, i.e.,
    \begin{equation}
        u(k) := u(k-1)+\Delta u^*_{0|k}.
        \label{eq:input-update}
    \end{equation}
    The gains are grouped as in~\eqref{eq:optimal-input-sequence} into $\hat{P}_1=KP_{1,I}$ and $\hat{P}_2=KP_{2,I}$. The implementation of iSPC~\eqref{eq:optimal-input-sequence}--\eqref{eq:input-update} is graphically shown in Fig.~\ref{fig:iDPC-Control-phase-controller}. As shown therein, only basic matrix vector multiplications are required, which can be executed by any industrial microprocessor.

        The parameters $P, \,Q, \,R, \,T, \,T_{\text{ini}}$ and $N$ are tuned as follows. A higher $P$ and $Q$ make the controller more aggressive as the tracking error penalty is higher. Increasing $R$ makes the controller less aggressive as faster-changing inputs are penalised. Increasing $T$ allows for more dynamics to be captured in $\Theta^*$, whereas increasing $T_{\text{ini}}$ gives a better initial condition estimation, both yielding a more accurate prediction. Increasing $N$ makes the controller less aggressive, as more long-term effects are taken into account, which tends to give a smoother response.
        
        As stated in~\cite{Lazar2022Offset-freeControl}, two general rules of thumb for linear systems are $N\geq T_{\text{ini}}\geq n_I:=n_u+n_y$ and $T\geq (T_{\text{ini}}+N)n_u + n_I$. This is, however, the case for linear systems.

    \subsection{Implementation}
        Two options for data collection are implemented and compared, namely fault-triggered iSPC (FT iSPC) and regular iSPC. The former means that the identification is done during the fault when oscillations are detected. For the latter, it holds that identification was done some time before the fault in nominal operation, and the controller is active during nominal conditions already, and remains active when the fault occurs without requiring fault-time data. This leads to a trade-off between performance and time to activation. The FT iSPC is closer to the fault dynamics but uses less data and has delayed activation. Regular iSPC captures different dynamics, but uses more data and is active from the start.  
        
\begin{figure}
            \centering
            \begin{subfigure}{\columnwidth}
                \centering
                \includegraphics[width=.8\columnwidth]{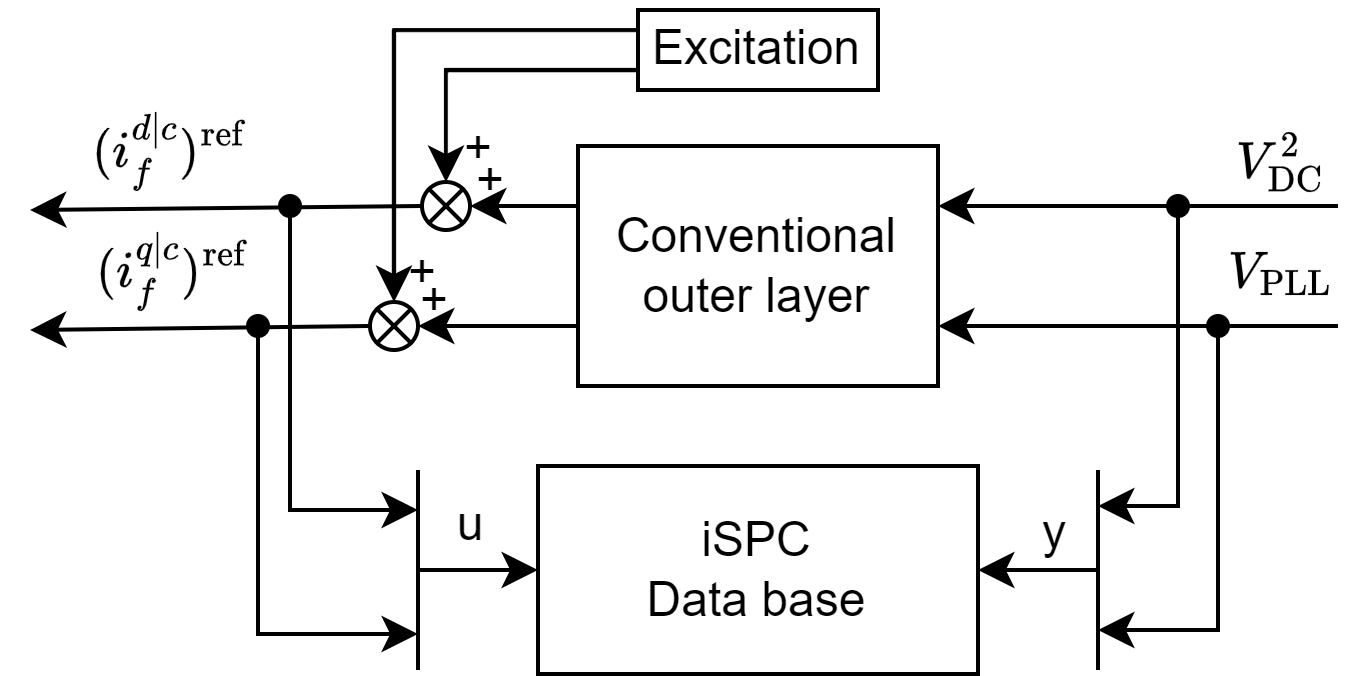}
                \subcaption{iSPC identification}
                \label{fig:iDPC-ID-phase}
            \end{subfigure}
        
            \begin{subfigure}{\columnwidth}
                \centering
                \includegraphics[width=.5\columnwidth]{iSPC_Control_phase_controller.png}
                \subcaption{iSPC control}
                \label{fig:iDPC-Control-phase}
            \end{subfigure}
            \caption{Controller structures during the two phases used in iSPC: (a) the identification phase (b) the control phase.}
            \label{fig:iSPC-controller}
        \end{figure}

        The steps necessary for the data collection and operation of iSPC are summarised in Algorithm~\ref{alg:online-iSPC}, and a graphic overview of the controller during identification and online control is shown in Fig.~\ref{fig:iSPC-controller}.
        \begin{algorithm}[t!]
            \caption{iSPC data-generation and operation}
            \label{alg:online-iSPC}
            \textbf{Input:} $(T,N,T_{\text{ini}}), (P,Q,R), r_y$\\
            \textbf{Identification:}
            \begin{algorithmic} [1] 
                \State Add excitation (e.g. white noise) to $u(k)=(I_{f}^{dq|c})^{\text{ref}}$
                \State Record $u(k)$ and $y(k)$
            \end{algorithmic}
            \textbf{Predictor calculation:}
            \begin{algorithmic} [1] 
                \State Construct Hankel matrices from \eqref{eq:hankel-matrices}
                \State Obtain $\Theta^*$ from \eqref{eq:theta-estimation} and extract $P_{1,I},P_{2,I}\text{ and }\Gamma_I$
            \end{algorithmic}
            \textbf{iSPC:}
            \begin{algorithmic} [1] 
                \For{$k = 0,1,2,\dots$}
                    \State Measure $y(k)$ 
                    \State Construct $\Delta\mathbf{u}_{\text{ini}}(k)$ and $\mathbf{y}_{\text{ini}}(k)$ from \eqref{eq:initial-io-trajectories}
                    \State Construct $\Delta \mathbf{u}^*(k)$ from \eqref{eq:optimal-input-sequence}
                    \State Apply $u(k)$ from \eqref{eq:input-update} and save it
                \EndFor
            \end{algorithmic}
        \end{algorithm}

\section{iSPC validation in weak grid faults}\label{sec:num-ex}
    The grid parameters and CC gains for the simulations were taken from~\cite{Li2019StabilityGrids} and are found in Table~\ref{tab:sim-res}. The initial conditions were determined through load-flow calculations assuming nominal grid and PCC voltage, and 90\% of nominal power export to the grid.
    \begin{table}
        \caption{Simulation Parameters for a GCPC}
        \centering
        \resizebox{\columnwidth}{!}{%
            \begin{tabular}{|c|c|c|c|}
                \hline
                Parameter & Symbol & Value & Unit \\ \hline
                Nom. frequency & $\omega_0$ & $2\pi\,60$ & rad/s \\ \hline
                Nom. conv. power & $P^{\text{base}}$ & 2 & MW \\ \hline
                Nom. conv. voltage & $V_{\text{co}}^{\text{base}}$ & 575 & V \\ \hline
                Nom. DC-link voltage & $(V_{\text{DC}}^2)^{\text{base}}$ & 1100 & V \\ \hline
                DC-link capacitor & $C_{\text{DC}}$ & 90 & mF \\ \hline
                Filter components 
                  & \makecell[c]{$L_f (X_f), R_f,$ \\ $C_f (B_f)$} 
                  & \makecell[c]{0.15, 0.003, \\  0.178} & \makecell[c]{p.u., p.u., \\ p.u.} \\ \hline
                \makecell[c]{Transmission cable\\components} 
                  & $L_{\text{ca}}(X_{\text{ca}}), R_{\text{ca}}$ 
                  & 0.45, 0.045 & p.u., p.u. \\ \hline
                Nom. grid parameters & $L_g(X_g)$ & 0.01 & p.u. \\ \hline
                Inner layer gains 
                  & $K_{i}^{p}, K_{i}^{i}$ & 0.48, 3.27 & -, - \\ \hline
                Conv. DC-link gains 
                  & $K_{P}^{p}, K_{P}^{i}$ & 0.4, 40 & -, - \\ \hline
                Conv. PCC voltage gains 
                  & $K_{v}^{p}, K_{v}^{i}$ & 0.25, 25 & -, - \\ \hline
                Conv. PLL gains 
                  & $K_{\text{PLL}}^{p}, K_{\text{PLL}}^{i}$ & 60, 1400 & -, - \\ \hline
            \end{tabular}%
        }
    \end{table}
    
    \subsection{Fault scenario}
        All fault simulations use similar timing. The controllers start with nominal conditions ($X_g = L_g\omega_0 = 0.01$ p.u. and $V_g = 1$ p.u.). At $t=1$ s, the fault occurs, which lasts till $t=5$ s, after which nominal conditions are restored. This fault models higher short-circuit ratios and grid voltage steps. The duration of the fault is chosen to demonstrate the required time for oscillation detection and fault-time data collection in the case of FT iSPC, rather than being close to realistic fault times.
    
        For the FT iSPC controller, similar to the DPC controllers in~\cite{Markovsky2023Data-DrivenSystems, Zhao2024DirectOptimization}, the system is given one second of oscillation detection before the identification and subsequent control phase start. Regular iSPC does not use any fault-time data. The iSPC controller parameters are given in Table~\ref{tab:con-param}. The controller parameters $N$, $Q$ and $R$ were tuned heuristically. $T_{ini}$ had to be at least larger than the system order, or the number of states of the system. $T_{FT iSPC}$ and $T_{iSPC}$ can be as large as desired, with more data leading to the multi-step predictor converging to the true linearised dynamics around the operating point, even in noisy situations due to the least-squares identification\cite{Lazar2022Offset-freeControl}. Due to the non-linear system, having more data allows a linear predictor to be fitted over a wider range of dynamics of the real system. $T_{FT iSPC}$ has to be as short as possible for fast controller activation. $T_s$ for the outer loop was chosen in line with earlier research on DPC for GCPC\cite{Zhao2024DirectOptimization, Markovsky2023Data-DrivenSystems}, though due to lower computation complexity, it could be decreased further for more accurate control.
        
        In all the results, timespans where the relevant controllers have settled are not shown to highlight the transient responses. For similar reasons, the time-axis scaling is changed.
        \begin{table}
            \caption{iSPC Control Parameters}
            \label{tab:con-param}
            \centering
            \renewcommand{\arraystretch}{1.3}
            \resizebox{\columnwidth}{!}{%
                \begin{tabular}{|c|c|c|c|}
                    \hline
                    Parameter & Symbol & Value & Unit \\ \hline
                    Controller sample time & $T_s$ & 1 & ms \\ \hline
                    \makecell[c]{Dataset size} & $T_{\text{FT\,iSPC}},T_{\text{iSPC}}$ & $750,\,10^4$  & - \\ \hline
                    \makecell[c]{Dataset parameters}
                      & $N, T_{\text{ini}}$ & 50, 25 & -, - \\ \hline
                    \makecell[c]{Cost function\\parameters}
                      & $Q(=P), R$ 
                      & \makecell[c]{%
                        $\displaystyle
                        \begin{matrix} 
                          8.5 \,\text{diag}(0.85,1.30)\\
                          120 \,\text{diag}(1.2,1)
                        \end{matrix}$} 
                      & -, - \\ \hline
                \end{tabular}%
            }
        \end{table}
        
    \subsection{Simulation Results}
\begin{figure}
            \centering
            \includegraphics[width=\columnwidth]{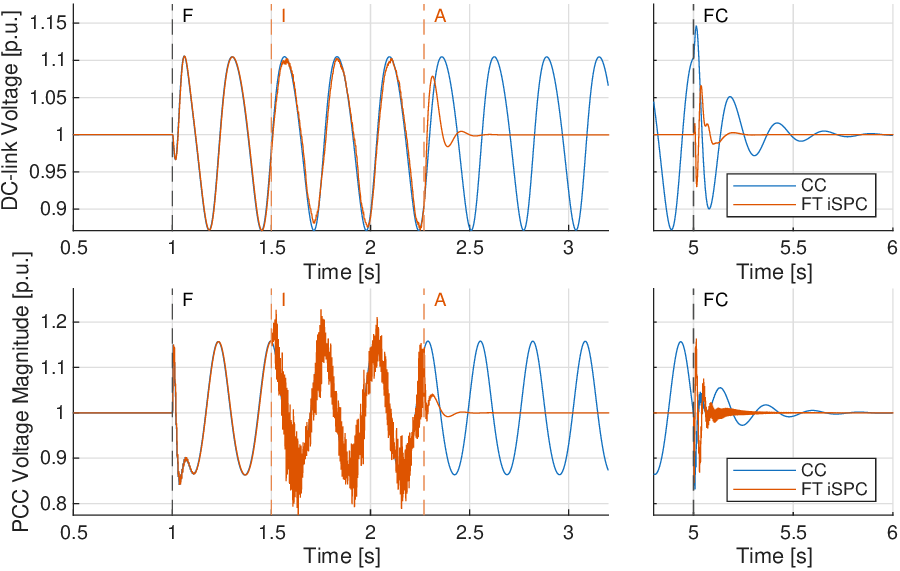}
            \caption{Simulated response of a GCPC under a critically weak grid fault: CC alone versus CC with iSPC after the fault occurs. F, FC, I and A indicate the start and the clearing of the fault and the start of the identification and the activation of FT iSPC, respectively.}
            \label{fig:GCPC response FT iSPC}
        \end{figure}

\begin{figure}
            \centering
            \includegraphics[width=\columnwidth]{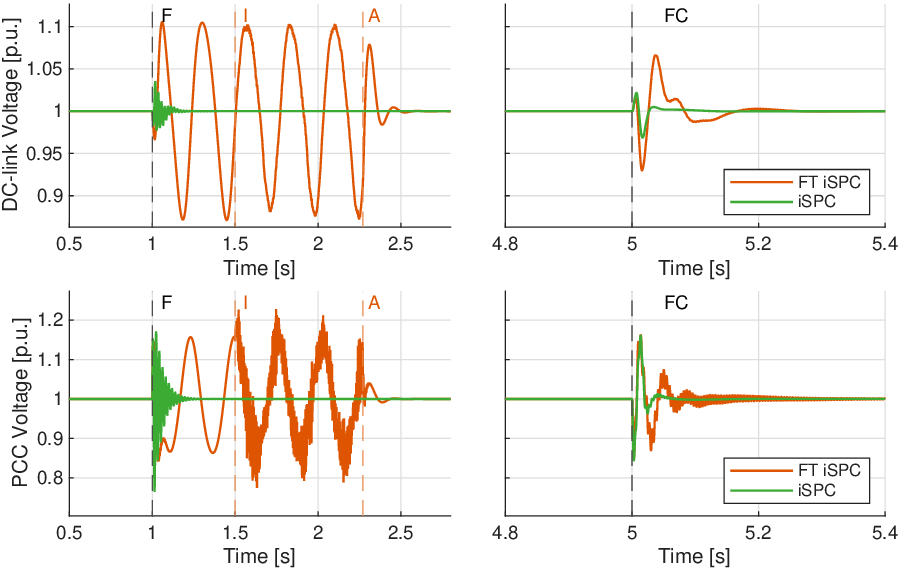}
            \caption{Simulated response of a GCPC under a critically weak grid fault: CC with FT iSPC after the fault versus regular iSPC. F, FC, I and A have the same meaning as in Fig.~\ref{fig:GCPC response FT iSPC}.}
            \label{fig:GCPC response iSPC}
        \end{figure}

        The first scenario simulates a critical fault and compares the response of CC with FT iSPC and FT iSPC with regular iSPC, respectively. In this scenario, the grid voltage drops to $V_g=0.96$ p.u. and the grid reactance jumps to $X_g = 0.1819$ p.u. This last value was determined such that CC shows sustained oscillations. In Fig.~\ref{fig:GCPC response FT iSPC}, the results of CC and FT iSPC are shown, and Fig.~\ref{fig:GCPC response iSPC} shows FT iSPC and regular iSPC. Therein, the dashed lines marked with F, FC, I and A indicate the moments the fault occurs and is cleared, and when the FT iSPC identification and activation phases start, respectively. Identification of the regular iSPC predictor is done under nominal conditions outside of the scenario, which allows regular iSPC to be active from the start. 

        To quantify controllers performance, two metrics are considered. The first, the settling time, is defined as the time for the outputs to reach and remain within a tube around the steady-state value. CC oscillates around that steady-state value with a certain amplitude. The radius of the tube is 2\% of that oscillation amplitude. The second metric, the root mean squared error (RMSE), is computed over a 4-second window following the fault onset and the clearance. These results for CC, FT iSPC and regular iSPC are summarised in Table~\ref{tab:sim-res}, showing that regular iSPC consistently outperforms the other two controllers in both metrics.
        \begin{table}
            \centering
            \caption{Simulation Results CC, FT iSPC and iSPC}
            \label{tab:sim-res}
            \renewcommand{\arraystretch}{1.3} 
            \resizebox{\columnwidth}{!}{%
            \begin{tabular}{|l|l|ll|ll|}
                \hline
                \multicolumn{2}{|c|}{} 
                  & \multicolumn{2}{c|}{Settling time (s)} 
                  & \multicolumn{2}{c|}{RMSE (p.u.)} \\ \cline{3-6}
                \multicolumn{2}{|c|}{} 
                  & \multicolumn{1}{c|}{\makecell{During\\[-2pt]fault}} 
                  & \multicolumn{1}{c|}{\makecell{After\\[-2pt]fault}} 
                  & \multicolumn{1}{c|}{\makecell{During\\[-2pt]fault}} 
                  & \multicolumn{1}{c|}{\makecell{After\\[-2pt]fault}} \\ \hline
                \multirow{2}{*}{CC}  & $V_{\text{DC}}$        
                  & \multicolumn{1}{l|}{-} &  0.80 & \multicolumn{1}{l|}{$8.15\times 10 ^{-2}$} & $1.98\times 10 ^{-2}$ \\ \cline{2-6}
                                         & $V_{\text{PLL}}$ 
                  & \multicolumn{1}{l|}{-} & 0.63 & \multicolumn{1}{l|}{$1.05\times 10 ^{-1}$} & $1.26\times 10 ^{-2}$ \\ \hline
                \multirow{2}{*}{FT iSPC}  & $V_{\text{DC}}$        
                  & \multicolumn{1}{l|}{1.48} &  0.22 & \multicolumn{1}{l|}{$4.52\times 10 ^{-2}$} & $6.09\times 10 ^{-3}$ \\ \cline{2-6}
                                         & $V_{\text{PLL}}$ 
                  & \multicolumn{1}{l|}{1.42} & 0.30 & \multicolumn{1}{l|}{$6.01\times 10 ^{-2}$} & $1.11\times 10 ^{-2}$ \\ \hline
                \multirow{2}{*}{iSPC} & $V_{\text{DC}}$        
                  & \multicolumn{1}{l|}{0.16} & 0.05 & \multicolumn{1}{l|}{$2.19\times 10 ^{-3}$} & $2.06\times 10 ^{-3}$ \\ \cline{2-6}
                                         & $V_{\text{PLL}}$ 
                  & \multicolumn{1}{l|}{0.21} & 0.05 & \multicolumn{1}{l|}{$1.04\times 10 ^{-2}$} & $7.45\times 10 ^{-3}$ \\ \hline
            \end{tabular}
            }
        \end{table}
    
\begin{figure}
            \centering
            \includegraphics[width=\columnwidth]{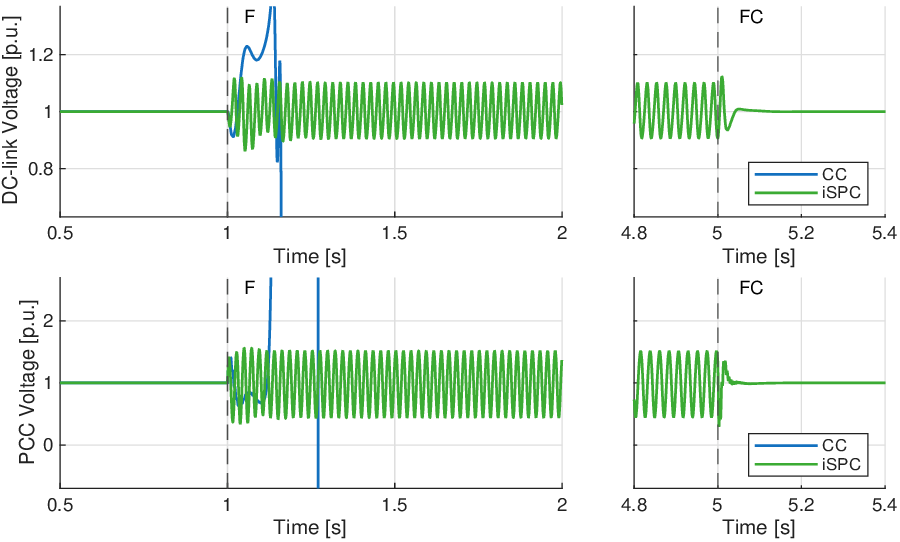}
            \caption{Simulated PCC voltage response of a GCPC under a critically weak-grid fault: CC versus regular iSPC. F and FC have the same meaning as in Fig.~\ref{fig:GCPC response FT iSPC}.}
            \label{fig:GCPC Improved Control Lim}
        \end{figure}

        Finally, the robustness of regular iSPC is assessed by increasing $X_g$ until it exhibits oscillatory behaviour after the fault. The stability is maintained till $X_g=0.3595$ p.u., which is almost double the critical value for which CC can maintain oscillatory but still (marginally) stable behaviour. Fig.~\ref{fig:GCPC Improved Control Lim} illustrates the response in this extreme case of both CC and regular iSPC. While CC becomes unstable very quickly, regular iSPC manages to maintain sustained oscillations despite the severe fault and only pre-fault data, and it recovers quickly after the fault is cleared.
        The computations were done in MATLAB/Simulink 2025a on a laptop (12th Gen Intel(R) Core(TM) i7-12700H, clocked at 2.30 GHz, 16 GB RAM, Windows 11). The Matlab function block that calculates the iSPC input took on average 0.242 ms, which is well within the 1 ms sampling time.
        
\section{Conclusion and recommendations}
    This paper highlighted the need for improved GFL GCPC control through the simulation of a GCPC during a critically weak grid fault. Next, two DPC controllers were tested: FT iSPC and regular iSPC.
    
    Regular iSPC outperformed both CC by settling the system, and did so by almost an order of magnitude faster than FT iSPC. It also reduced the RMSE by a factor of 40 compared to CC, and by a factor of two compared to FT iSPC. This demonstrates that a simple analytical iSPC controller, identified pre-fault under nominal conditions, can already substantially improve GCPC control and, given the time required for calculation, be implemented in standard GCPC microprocessor hardware.
    
    Finally, the limits of the iSPC controller with pre-fault data were determined, leading to a doubling of the critical equivalent grid reactance for a 4\% grid voltage drop.

    Future work will include verifying the controller with constraints, testing an adaptive version of iSPC, and investigating the controller response with more advanced grid models, such as the IEEE 9-bus or 4-machine 2-area systems~\cite{Vittal2012PowerControl}, or the CIGRÉ MV Benchmark System used for distribution network tests.
    
\bibliographystyle{IEEEtran}
\bibliography{references}

\end{document}